\documentstyle[eqsecnum,aps,epsf]{revtex} 
\newcommand{\postscript}[2] {\setlength{\epsfxsize}{#2\hsize}
\centerline{\epsfbox{#1}}}

\begin{document}
\twocolumn[\hsize\textwidth\columnwidth\hsize\csname@twocolumnfalse\endcsname

\title{Numerical solution of the two-dimensional
 Gross-Pitaevskii equation for trapped  interacting atoms}

\author{Sadhan K. Adhikari}
\address{Instituto de F\'{\i}sica Te\'orica, Universidade Estadual
Paulista, 01.405-900 S\~ao Paulo, S\~ao Paulo, Brazil\\}

\date{\today}
\maketitle
\begin{abstract}

We present a numerical scheme for solving the time-independent nonlinear
Gross-Pitaevskii equation in two dimensions describing the Bose-Einstein
condensate of trapped interacting neutral atoms at zero temperature.  The
trap potential is taken to be of the harmonic-oscillator type and the
interaction both attractive and repulsive.  The Gross-Pitaevskii equation
is numerically integrated consistent with the correct boundary conditions
at the origin and in the asymptotic region.  Rapid convergence is obtained
in all cases studied. In the attractive case there is a limit to the
maximum number of atoms in the condensate.

{\bf Accepted in Physics Letters A}
\end{abstract} 

\vskip1.5pc] 
Recently, there have been experiments \cite{1} of Bose-Einstein
condensation in dilute bosonic atoms (alkali and hydrogen atoms) employing
magnetic traps at ultra-low temperatures. These experiments have
intensified theoretical investigations on various aspects of the
condensate \cite{2,3,4,5,6,7}.  The condensate can consist of few thousand
to millions of atoms confined by the trap potential. The properties of the
condensate at zero temperature are usually described by the nonlinear
time-independent mean-field Gross-Pitaevskii (GP)  equation \cite{8},
which properly incorporates the trap potential as well as the interaction
among the atoms.  The effect of the interaction leads to a nonlinear term
in the GP equation which complicates its solution procedure.

There have been a series of recent studies which deal with the numerical
solution of the three-dimensional GP equation \cite{3,4,5,6}. These works
have emphasized the serious difficulties in obtaining numerical
convergence of the solution.  There has been no such systematic study on
the numerical solution of the GP equation in two dimensions. Although,
there has been no experiments on Bose-Einstein condensation of
two-dimensional systems, this is a problem of great interest.  A system of
ideal Bose gas in two dimensions does not undergo condensation at a finite
temperature \cite{9}. However, condensation can take place under the
action of a trap potential \cite{10}. In order to achieve two-dimensional
Bose-Einstein condensation in real three-dimensional traps, one should
choose the frequency $\omega_{xy}$ in the $x-y$ plane negligibly small
compared to that in the $z$ direction $\omega_z$ \cite{11}. Also, there
has been consideration of Bose-Einstein condensation in low-dimensional
systems for particles confined by gravitational field or by a rotational
container \cite{12}. Possible experimental configurations for
Bose-Einstein condensation in spin-polarized hydrogen in two dimensions
are currently being discussed \cite{2,7}. Because of these interests, here
we perform a critical study of the numerical solution of the
time-independent GP equation in two dimensions for an interacting Bose gas
under the action of a harmonic-oscillator-type trap potential.  The
interatomic interaction is taken to be both attractive and repulsive in
nature.

In steady state at zero temperature the condensate wave function is
described by the following effective nonlinear Schr\"odinger-type equation
known as the Gross-Pitaevskii equation \cite{8} for condensed neutral
bosons in a harmonic trap:  \begin{eqnarray}\label{1} \left[
-\frac{\hbar^2}{2m}\nabla^2 +\frac{1}{2}m\omega^2 r^2 +g\Psi^2({\bf
r})-\mu
\right]\Psi({\bf r})=0.  \end{eqnarray} Here $\Psi({\bf r})$ is the
condensate wave function at position ${\bf r}$, $m$ the mass of a single
bosonic atom, $ m\omega^2 r^2/2$ the attractive
harmonic-oscillator trap potential, $\omega$ the oscillator frequency,
$\mu$ the chemical potential and $g$ the strength of interatomic
interaction. A positive $g$ correspond to a repulsive interaction and a
negative $g$ to an attractive interaction. 

Here we shall be interested in the spherically symmetric solution
$\Psi({\bf r})\equiv \psi({ r})$ to  eq. (\ref{1}) which can be written as
\begin{eqnarray}\label{2} \left[
-\frac{\hbar^2}{2m}\frac{1}{r}\frac{d}{dr}r\frac{d}{dr}
+\frac{1}{2}m\omega^2 r^2 +g\psi^2({ r})-\mu \right]\psi({ r})=0. 
\end{eqnarray} The ground state of the condensate appears in such a
spherically symmetric state. As is Ref. \cite{6}, it is convenient to
express  eq. (\ref{2})  in terms of dimensionless variables defined by $x
=
r/a$, where $a\equiv \sqrt {\hbar/(m\omega)}$, $\alpha = \mu/(\hbar
\omega)$, $ \psi(x) = a\sqrt{2mg}\psi(r)/\hbar$. In terms of these
dimensionless variables  eq. (\ref{2}) can be written as
\begin{eqnarray}\label{3} \left[ -\frac{1}{x}\frac{d}{dx}x\frac{d}{dx}
+x^2 +c\psi^2({x})-2\alpha \right]\psi({ x})=0.  \end{eqnarray} where
$c=\pm 1$ carries the sign of $g$, $c=1$ corresponds to a repulsive
 interaction and $c=-1$ corresponds to an attractive  
interaction.

The normalization of the wave function is given by 
\begin{equation}\label{4}
N= 2\pi \int_0^\infty dr r \psi^2(r) ,
\end{equation}
where $N$ is the total number of atoms in the condensate. In terms of the
dimensionless variables defined above, this normalization condition
becomes
\begin{equation}\label{5}
 \int_0^\infty  x dx \psi^2(x)=n \equiv \eta N, 
\end{equation}
where as in Ref. \cite{2} we have introduced a dimensionless coupling 
$\eta \equiv m g /(\pi \hbar^2)$ of interatomic interactions and a
reduced number of particles $n$. 
From a study of the temperature dependence of the chemical potential for 
a system of interacting bosons  in two dimensions it
was concluded in Ref. \cite{2} that a Bose-Einstein-condensate-like
behavior is
obtained in this system for  values of $\eta$ typically smaller than
0.001. Hence for a qualitative estimate of the number of condensed bosons
in this calculation,
we
shall consider  $\eta =0.0001$.

Instead of solving eqs. (\ref{2})  and (\ref{4}), we shall be working with
eqs.  (\ref{3})  and (\ref{5}). Equation (\ref{3}) is independent of all
parameters of the problem, such as, $m$, $\omega$, $g$, and $N$. The
relevant parameters appear in the normalization condition (\ref{5}). The
constant $n$ in eq. (\ref{5}) is the reduced number of atoms for the
system and is related to the real number of atoms $N$. 

Another interesting property of the condensate wave function is its
mean-square radius defined by
\begin{equation}\label{6}
\langle r^2 \rangle = \frac{2\pi}{N}\int_0^\infty r^2 \psi^2(r) rdr. 
\end{equation}
In terms of the dimensionless variables defined above we have 
\begin{equation}\label{7}
\langle x^2 \rangle = \frac{1}{n}\int_0 ^\infty  x^2 \psi^2(x) xdx. 
\end{equation}

To solve  eq. (\ref{3}) numerically, first we study the asymptotic
behavior
of its solutions. Since for a sufficiently large $x$, $\psi(x)$ must
vanish asymptotically, the nonlinear term proportional to $\psi^3(x)$
can eventually be neglected in  eq. (\ref{3}) in the asymptotic region.
Thus for large  $x$, this equation can be approximated by 
\begin{eqnarray}\label{8}
\left[ -\frac{1}{x}\frac{d}{dx}x\frac{d}{dx} +x^2
-2\alpha \right]\psi({ x})=0.
\end{eqnarray}
If this equation would be valid for all $x$ this would be 
the equation for the two-dimensional oscillator in the spherically
symmetric state
permitting solution for $\alpha = 1, 3, 5, ...$ etc. However, in the
present problem  eq. (\ref{8}) is valid only in the asymptotic region.
Considering  eq. (\ref{8})  as a mathematical equation valid for all
$\alpha$, the asymptotic form of the physically acceptable solution is
given by 
\begin{equation}\label{9}
\lim_{x \to \infty} \psi(x)= N_C\exp \biggr[-\frac{x^2}{2}+(\alpha -
1)\ln x
\biggr],   
\end{equation}
where $N_C$ is a normalization constant. The derivative of the wave
function
in the asymptotic region can be obtained from  eq. (\ref{9}) and 
one obtains the following asymptotic form  for the log-derivative  of the
wave function 
\begin{equation}\label{10}
\lim_{x \to \infty} \frac {\psi'(x)}{\psi{(x)}}=  \biggr[-x+
\frac{\alpha-1}{x}\biggr],
\end{equation}
which is independent of the normalization constant $N_C$ of  eq. (\ref{9})
and where the prime denotes derivative with respect to $x$.

Next we consider  eq. (\ref{3}) as $x\to 0$. The nonlinear term approaches
a constant in this limit because of the regularity of the wave function at
$x=0$. Then one has the following usual  conditions 
\begin{equation}
\psi(0)={ \mbox{constant}}, \hskip 1cm 
\psi'(0)=0 , \label{11}
\end{equation}
 as in the case of the two-dimensional harmonic oscillator problem.

Equation (\ref{3}) is integrated numerically for a given $\alpha$ by the
four-point Runge-Kutta rule starting at the origin ($x=0$) with the
initial boundary condition (\ref{11}) with a trial $\psi(0)$. The
numerical integration is performed in steps of $dx = \Delta$, where
$\Delta $ is typically taken to be  0.0001. The
integration is propagated to $x = x_{\mbox {max}}$, where the asymptotic
condition
(\ref{10}) is valid. The agreement between the numerically calculated
log-derivative of the wave function and the theoretical result (\ref{10})
was enforced to five significant figures. 
The maximum value of $x$, up to
which we needed to integrate  (\ref{3}) numerically for obtaining this
precision, is $x_{\mbox {max}}= 5 $. If for a trial $\psi(0)$, the
agreement of the log-derivative can not be obtained, a new value of
$\psi(0)$
is
to be chosen.  The proper choice of $\psi(0)$ was implemented by the
secant  method. Even with this method, sometimes it is
difficult to obtain the proper value of $\psi(0)$ for a given $\alpha$.
Unless
the initial guess is ``right"  and one is sufficiently near the desired
solution the method could fail and lead to either the trivial solution
$\psi(x) = 0$ or an exponentially divergent nonnormalizable solution in
the asymptotic region. However,
there is one guideline which is of help in finding the proper value of
$\psi(0)$.
The convergence of the numerical log-derivative to the theoretical result
(\ref{10}) with a variation of $\psi(0)$ for a fixed $\alpha$ is monotonic
in
nature provided that one is near the exact value. If for two trial values
of $\psi(0)$ near the exact value one obtains a positive and a negative
error
for the log-derivative, respectively, the exact $\psi(0)$ lies in between
these
two trial values.

For both attractive ($c= -1$) and repulsive ($c=1$) interatomic
interactions, in addition to the ground state solution with no nodes, 
eq. (\ref{3}) also permits radially excited states with nodes in the wave
function.  In the absence of the nonlinear term, the spherically-symmetric
discrete ground-state solution of the two-dimensional oscillator given by
eq.  (\ref{3}) is obtained for $\alpha = 1$. Radially excited oscillator
states appear for $\alpha =3, 5, ...$ etc. In the presence of the
nonlinearity, for attractive interatomic interaction ($c=-1$), the
solutions of the GP equation for the ground state appear for values
chemical potential $\alpha <1$.  For a repulsive interatomic interaction
($c=1$), these solutions appear for $\alpha > 1$. The solutions of eq. 
(\ref{3}) for the radially excited state with one node for the attractive
($c=-1$) and repulsive ($c=1$) cases appear for values of chemical
potential $\alpha$ smaller and larger than the energy of the first excited
state ($\alpha = 3$) of the harmonic oscillator problem, respectively.

\vskip -3.5cm
\postscript{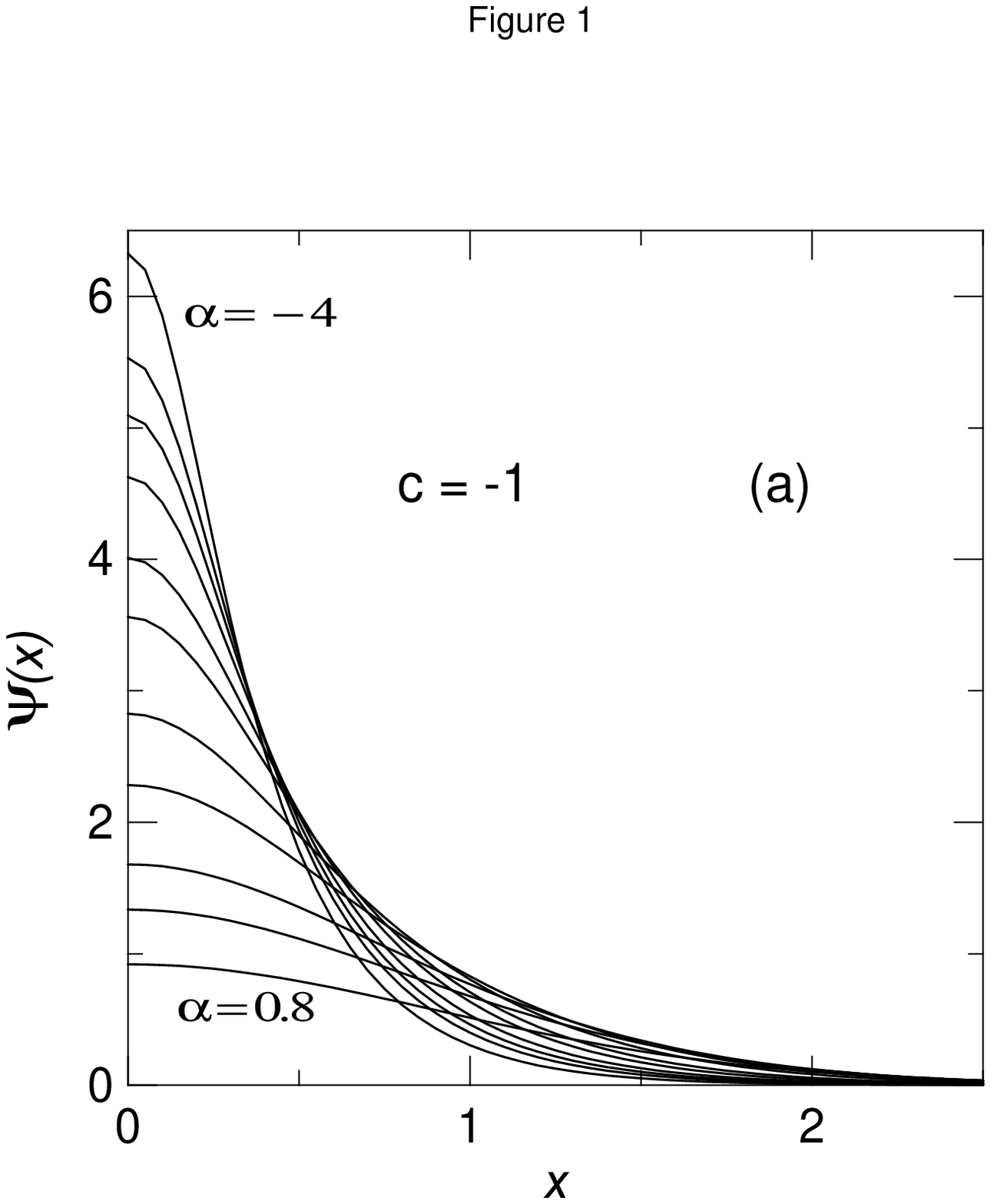}{1.0}    
\vskip -2.0cm
\vskip -3.4cm
\postscript{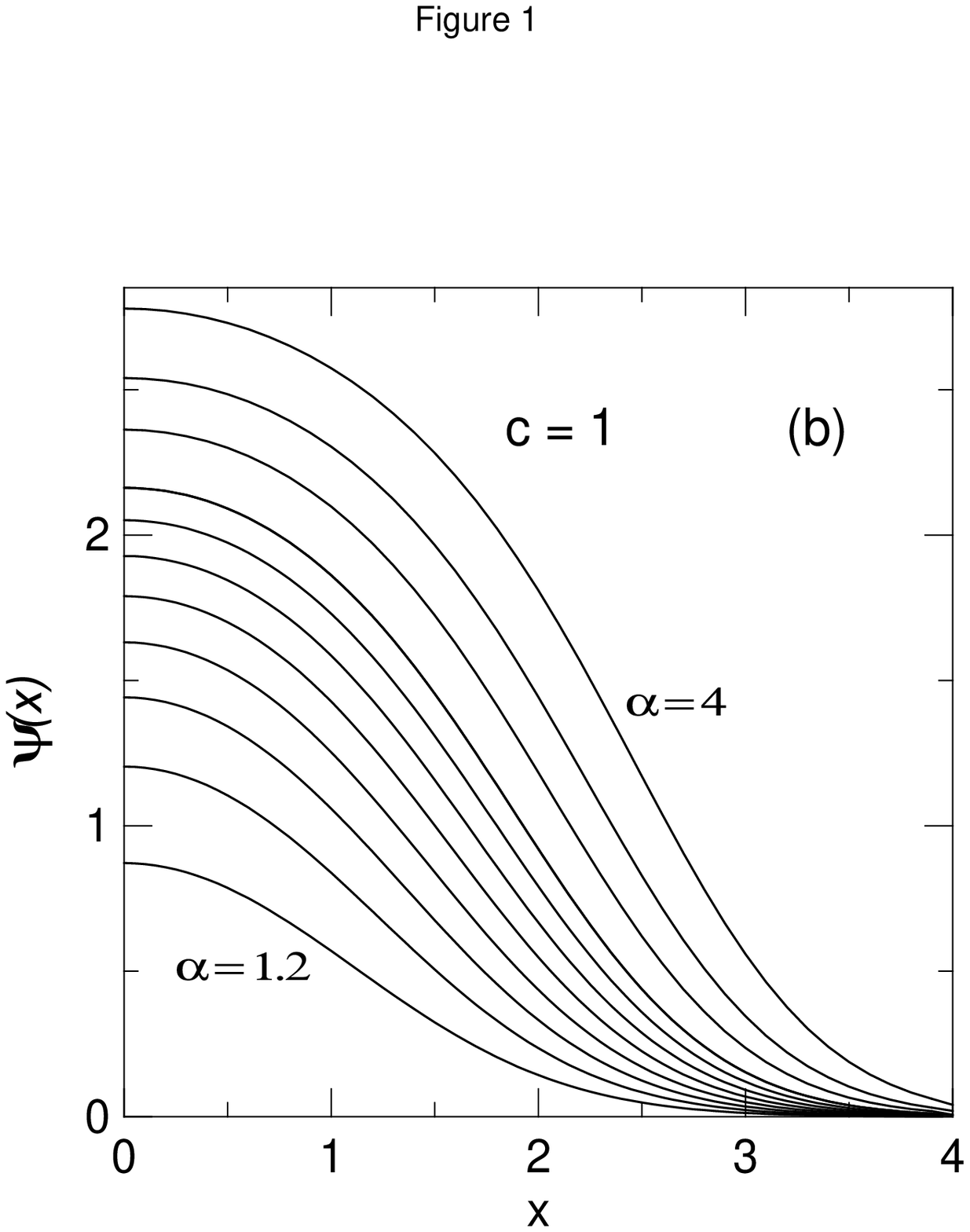}{1.0}    
\vskip -2.1cm

\vskip .9cm

{ {\bf Fig. 1.}
 Ground-state condensate wave function $\psi(x)$ versus $x$ for (a)
attractive and
(b) repulsive interparticle interactions.  The parameters for these
cases are given in Table I. The curves appear in the same order as in
Table I. The lowermost curve corresponds to the first row in Table I.}

\vskip .2cm

First we consider the ground-state solution of eq. (\ref{3}) for
different $\alpha $ in the cases of both  attractive and repulsive
interactions. 
The relevant parameters for these
solutions (values of the wave-function at the origin
$\psi(0)$,  reduced number $n$, and mean-square radii $\langle x^2
\rangle$)
are listed in table 1. 
The wave functions for different values of $\alpha$ for the
attractive and repulsive interparticle interactions for the cases shown in 
table 1 are exhibited  in figures 1(a) and 1(b), respectively, where we
plot
$\psi(x)$ versus $x$. The curves in figures 1(a) and
1(b) appear in the same order as the rows in table 1 and it is easy to
identify
the corresponding values of $\alpha$ from  the values of $\psi(0)$
 of each curve. 
From figures 1(a) and (b) we find
that the nature of the wave function for these two cases are quite
different.

{Table 1: Parameters for the numerical solution of the GP equation
(\ref{3}) for $c=\pm 1$ for the ground state wave function. The first four
columns refer to the attractive interaction $c=-1$ and the last four
columns refer to the repulsive interaction  $c=1$. 
}
\vskip
.2cm

{\begin{center}{\begin{tabular} {|c|c|c|c|c|c|c|c|}
\hline
$\alpha$  & $\psi(0)$   & $n$  & $\langle x^2\rangle$ & 
$\alpha$  & $\psi(0)$   & $n$  & $\langle x^2\rangle$ \\
\hline
1.0 & 0    & 0     &  0      &   1.0  &0 &0  & 0 \\
0.8 & 0.9185    & 0.3663     &  0.9030   &1.2  & 0.8719& 0.4353&1.1027 
\\
0.6 & 1.3347    & 0.6690     &  0.8127  &  1.4 &1.2036 &0.9435 &1.2103 \\
0.4 & 1.6795    & 0.9147     &  0.7297       &  1.6 &1.4415 &1.5276
&1.3219\\
          0.0 & 2.2827    & 1.2655    &  0.5872      &1.8   &1.6308
&2.1894
&1.4368 \\
   $-$0.4 & 2.8255    & 1.4798     &  0.4757     & 2.0  &1.7896 &2.9303
&1.5544\\
$-$1.0 & 3.5624    & 1.6530     &  0.3564    & 2.2  &1.9276 &3.7509
&1.6741\\
$-$1.4 & 4.0097    &1.7152      &  0.3005       & 2.4  &2.0507 &4.6518
&1.7956 \\
$-$2.0 & 4.6249    &  1.7695    &  0.2400       & 2.6  & 2.1627&5.6331
&1.9186 \\
$-$2.5 &5.0942&1.7957&0.2040           & 3.0  & 2.3626&7.8377
&2.1679 \\
$-$3.0 &5.5307&1.8126&0.1767   & 3.4  & 2.5401&  10.3651
&2.4206 \\
$-$4.0&6.3252&1.8319&0.1385     & 4.0 & 2.7786&  14.7609
&2.8041 \\
\hline
\end{tabular}
}\end{center}}
\vskip .2cm

\vskip -3.2cm
\postscript{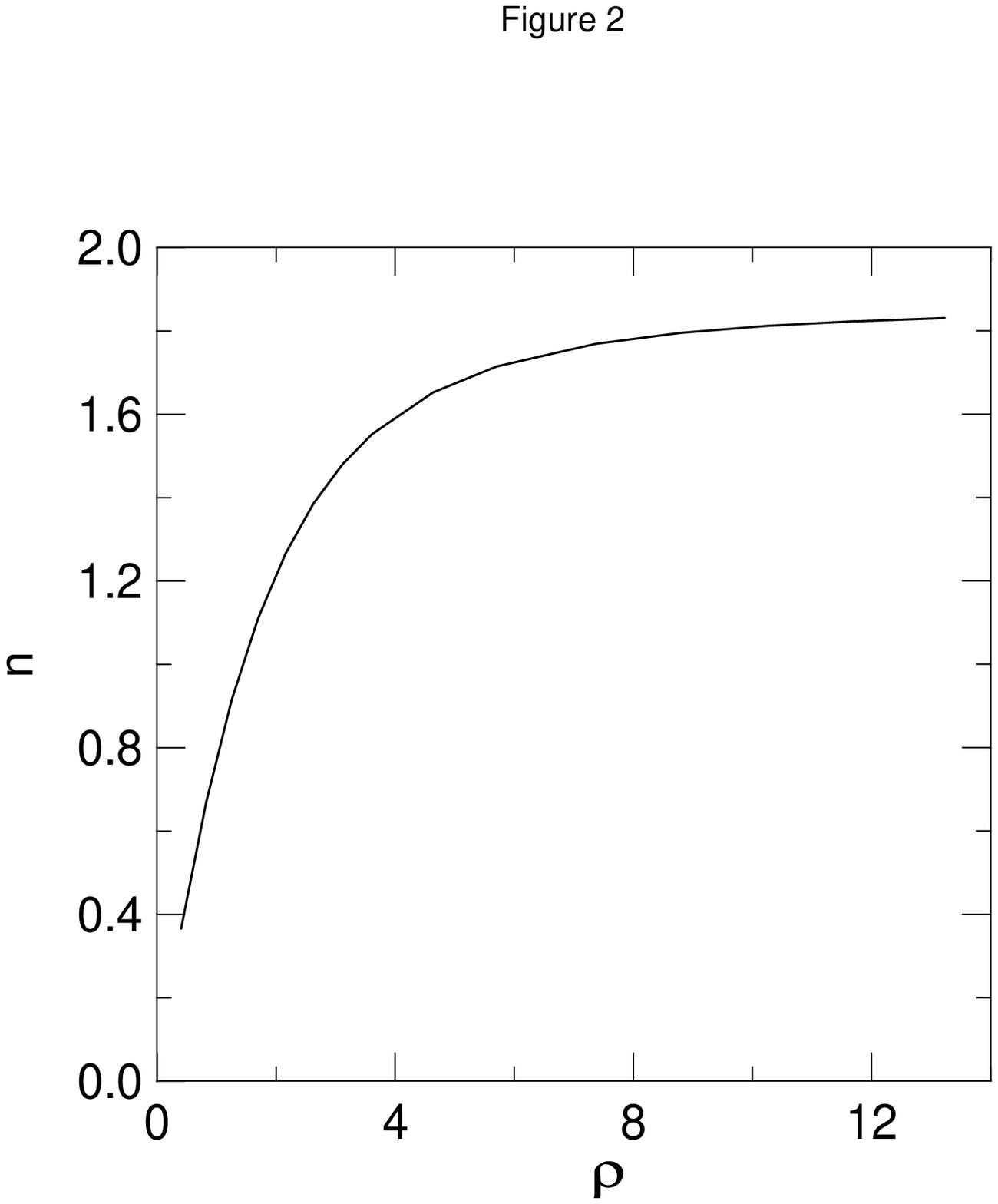}{1.0}    
\vskip -2.0cm

\vskip .2cm

{ {\bf Fig. 2.} The reduced number of particles $n$ versus density $\rho
\equiv
n/\langle x^2 \rangle$ of the ground-state condensate  for the attractive
interaction.}

{Table 2: Parameters for the numerical solution of the GP equation
(\ref{3}) for $c=\pm 1$ for the  wave function of the first excited state.
The first four
columns refer to the attractive interaction $c=-1$ and the last four
columns refer to the repulsive interaction  $c=1$. 
}
\vskip
.2cm

{\begin{center}{\begin{tabular} {|c|c|c|c|c|c|c|c|}
\hline
$\alpha$  & $\psi(0)$   & $n$  & $\langle x^2\rangle$ & 
$\alpha$  & $\psi(0)$   & $n$  & $\langle x^2\rangle$ \\
\hline
3.0 & 0    & 0     &  0      &   3.0  &0 &0  & 0 \\
2.9 & 0.9071  &  0.3940    &   2.9505      & 3.1  &  0.8814 &0.4064
&3.0505 \\
2.8 & 1.3012    & 0.7769    &  2.9012       & 3.2 & 1.2278 &0.8254
&  3.1013 \\
2.6 &1.8898     & 1.5118     &   2.8039      & 3.6  & 2.0006& 2.6466
&3.3101 \\
2.0  & 3.1912   & 3.5069     &  2.5215             & 4.0 &
2.4347& 4.7325
& 3.5282 \\
1.5  & 4.0643   &  4.9560    & 2.2976        & 4.5 &  2.7895& 7.7573
& 3.8124 \\
\hline
\end{tabular}
}\end{center}}
\vskip 0.2cm

For the attractive interparticle interaction, the wave function
is more sharply peaked at $x=0$ than in the case of the repulsive
interparticle interaction. 
Consequently, in the attractive case one has  
smaller values of the  reduced number 
$n$ and mean square radii $\langle x^2 \rangle$
 of the wave function. For the attractive case we find from table 1 that
with a reduction of the chemical potential $\alpha$
the reduced number   $n$ increases slowly and  the mean square radius
$\langle x^2 \rangle$ decreases rapidly, so that the density of the
condensate $\rho \equiv n/\langle x^2 \rangle$ tends to diverge as $n$ 
tends to a maximum value $n_{\mbox{max}}$. This means that 
there is  a maximum number of particles in the condensate in this
case. This peculiar
behavior in the attractive  case is demonstrated in figure 2 where we plot 
the reduced number $n$ of the condensate in the ground state versus 
the density $\rho$. From figure 2 we find that with the reduction of the
chemical potential, the reduced number of particles $n$ in the condensate 
 attains a saturation value. Numerically, we find this value to be
\begin{equation}\label{12}
n_{\mbox{max}}\equiv \eta N_{\mbox{max}} \approx 1.85.
\end{equation} 
However, during this process the mean
square radius continues to decrease thus leading to a divergence of the 
density of the condensate.  For $n>n_{\mbox{max}}$, there is no stable 
solution of the GP equation. 
There is no such limit on $n$ in the repulsive  case. In that case with
the increase of the chemical potential $\alpha$ the condensate 
increases in size as the number of particles in the condensate increases. 
These behaviors of the Bose-Einstein condensate in two dimensions were
also noted in three dimensions \cite{11,13}.

Next we consider solutions to  eq. (\ref{3}) with radial excitation. 
The present numerical method is equally applicable to the ground and all
radially excited states of the condensate.
In this work we shall  consider excited solutions
with only one node  for both attractive and repulsive interparticle
interactions, which we  discuss next. In table 2 we exhibit the
parameters of some of such solutions. In figures 3(a) and 3(b) we plot the
respective wave functions $\psi(x)$ versus $x$ for the attractive and
repulsive interactions. 
 We again find that in the case of attractive interaction 
the wave function is narrowly peaked near $x=0$, and for repulsive
interaction it is more extended in space to larger values of $x$.
Consequently, the reduced number  $n$ and mean square radii $\langle x^2
\rangle$ are larger in the case of repulsive interaction. We again find
from table 2 that in the attractive case, with a reduction of the
chemical potential $\alpha$, the mean square radius decreases as the
reduced number $n$ increases. It is expected that there should be a
saturation on $n$ in this case also. However, we did not try to find this
saturation numerically, which should occur for values of $\alpha$ much
smaller than those presented in table 2 leading to a much larger 
value of $n_{\mbox{max}}$ than in  eq. (\ref{12}).

Although, we have considered the problem in a system of dimensionless
units, it is interesting to see what our results correspond to in actual
units.  If we consider the typical value 0.0001 of $\eta$ favorable to the
formation of Bose-Einstein condensate \cite{2} as commented after  eq.
(\ref{5}), we can calculate the actual numer of atoms $N$.  The number of
atoms $N$ in the condensate for all cases reported in tables 1 and 2 is
maximum for the ground state of the repulsive interatomic interaction for
$\alpha =4$.  This number in this case is $N=147,600$. If we consider a
trap frequency such that $\sqrt{\hbar/(m \omega)} = 10,000$ \AA \cite{5},
then the root-mean-square radius for the above condensate with 147,600
atoms is 16,700 \AA.  
In the attractive case the
maximum numer of particles given by  eq. (\ref{12})  becomes 
$N_{\mbox{max}}= 18,500$.
Both the size and  numbers of particles seem to be
very reasonable for the condensate \cite{3,5}.

In this work we have investigated the numerical solution of the
Gross-Pitaevskii equation (\ref{1}) for Bose-Einstein condensation in two
dimensions under the action of a harmonic oscillator trap potential for
bosonic atoms interacting via both attractive and repulsive interparticle
interactions. In both cases we considered the wave function for the ground
state and radially excited state with one node. We expressed the GP
equation in dimensionless units independent of all parameters, such as,
atomic mass, harmonic oscillator frequency, number of atoms in the
condensate, and strength of atomic interaction. The relevant parameters
appear in the normalization condition (\ref{5}) of the wave function. We
derive the
boundary conditions (\ref{10}) and (\ref{11}) of the solution of the
dimensionless GP
equation (\ref{3}), which is integrated from the origin outwards in steps
of 0.0001 by the four-point Runge-Kutta rule consistent with the
boundary condition. At a particular value of the chemical potential, the
correct solution is obtained after a proper  guess of the boundary
condition of the wave function at the origin. From a initial trial value
of the wave function at the origin, Newton-Raphson method is used to
obtain the correct wave function after a matching with the boundary
condition in the asymptotic region. In both the ground and excited states
it is found that the wave function is sharply peaked near the origin for
attractive interatomic interaction. For a repulsive interatomic
interaction the wave function extends over a larger region of space.
In the case of an attractive potential,  the mean square radius decreases
with an increase of the number of particles in the condensate.
Consequently, 
a stable
solution of the GP equation can be obtained 
for a maximum number of particles in the condensate. For the ground state 
the maximum reduced number of particles is given by  eq. (\ref{12}).
 For the repulsive case there is no such limit on the number of particles
in the condensate.

We thank the authors of Ref. \cite{6} for a copy of their work 
prior to  publication.
The work is supported in part by the Conselho Nacional de Desenvolvimento
Cient\'\i fico e Tecnol\'ogico and Funda\c c\~ao de Amparo
\`a Pesquisa do Estado de S\~ao Paulo of Brazil.

\end{document}